\documentclass[aip,apr,reprint]{revtex4-1}

\usepackage{graphicx}
\usepackage{amssymb}
\usepackage{amsfonts}
\usepackage{amsmath}
\usepackage{amsbsy}
\usepackage{xcolor}
\usepackage[colorlinks=true, urlcolor=blue, citecolor=blue, linkcolor=blue]{hyperref}

\renewcommand{\vec}[1]{\mbox{\boldmath$\mathrm{#1}$}}
\let\sb=_ \catcode`\_=\active \def_#1{\ensuremath \sb{\rm#1}}

\begin{document}

\title{Non-destructive Ultrafast Steering of Magnetic Vortex by Terahertz  Pulses}

\author{Dongxing Yu}
\author{Jiyong Kang}
\affiliation{Key Laboratory for Magnetism and Magnetic Materials of MOE, Lanzhou University, 730000 Lanzhou, China}

\author{Jamal Berakdar}
\email{jamal.berakdar@physik.uni-halle.de}
\affiliation{Institut f\"ur Physik, Martin-Luther Universit\"at Halle-Wittenberg, 06099 Halle (Saale), Germany}

\author{Chenglong Jia}
\email{cljia@lzu.edu.cn}

\affiliation{Key Laboratory for Magnetism and Magnetic Materials of MOE, Lanzhou University, 730000 Lanzhou, China}

\begin{abstract}
\textbf{Electric control of  magnetic vortex dynamics in a reproducible way and on an ultrafast time scale is key element in the quest for efficient spintronic devices with low-energy consumption. To be useful the control scheme should ideally be swift, scalable, non-invasive,  and  resulting in reliable magnetic switching. Such requirements and in particular the reproducibility of  altering the vortex chirality and/or polarity are not yet met by magnetic vortex switching via external magnetic fields, spin-polarized currents, spin waves, or  laser pulses. Here we demonstrate a novel packaged-skyrmion mediated vortex switching process driven by a simple sequence of picosecond electrical field pulses via magneto-electric interactions. Both the vortex chirality and polarity show a well-defined reversal behaviour. The unambiguously repeated switching between four different magnetic vortex states provides an energy-efficient, highly localized and coherent control method for non-volatile magnetic vortex-based information storage and handling.}
\end{abstract}

\date{\today}
\maketitle

\section{Introduction}

Magnetic vortices are formed in confined soft ferromagnetic structures like disks, triangles, squares and stripes in the micro/sub-micron scales  \cite{Shinjo:2000,Wachowiak:2002}. The driving force is a subtle competition between geometrical balance of exchange interaction and shape anisotropy \cite{Antos:2008kb}. As one kind of topological magnetic defect \cite{Book:1998,Braun:2012}, a magnetic vortex is characterized by an in-plane curling magnetization (chirality: either clockwise or counterclockwise), and an out-of-plane nanometer-sized core magnetization (polarity: up or down). Having the simplest topologically non-trivial spin configuration,
magnetic vortices  exhibit  several advantageous features such as high thermal stability, a negligible magnetostatic interaction, and the diversity of information storage for high-density and nonvolatile magnetic memories and spintronic devices \cite{Bader:2006de,R. P. Cowburn:2007,B. Pigeau:2010,S. Sugimoto:2011}. 
Due to rotational symmetry of magnetic vortex and strong exchange interaction in ferromagnetic materials \cite{NA:2002}, it is however difficult to reverse the vortex polarity and/or chirality. Alternating magnetic fields \cite{B. Van:2006}, in-plane magnetic field pulses 
\cite{R. Hertel:2007}, resonant microwave pulses \cite{Pigeau:2010fy}, field-driven spin waves \cite{V. P. Kravchuk:2009,M. Kammerer:2011}, spin-polarized currents  \cite{K. Yamada:2007,VS:2007}, and photo-thermal-assisted femtosecond laser pulse excitation \cite{Anonymous:2018} have been used to switch the vortex-core polarity. To solve the problem of chirality switching with magnetic fields, structures with broken symmetry were also introduced \cite{J. Tobik:2012}. While few studies considered the simultaneous control of the vortex chirality and polarity,  it turned out to be challenging to precisely determine when the core switching occurs, and even worse, the clockwise and counterclockwise vortex states may emerge randomly with a similar occurrence frequency \cite{Anonymous:2018}, thus largely prohibiting the development of reliable  magnetic vortex-based spintronics.
%
%

To remedy this situation and facilitate the vortex dynamics for ultrafast all-optical magnetism  \cite{Kimel:2019}, we exploit in this work another feature of non-collinear spin ordering, namely  the
magnetoelectric (ME) effect \cite{Eerenstein:2006km,Simpson:2006uw,Tokura:2014eg,Zheng:2017bq} leading to a spin-driven emergent ferroelectric polarization $\vec{P}$ which  allows an external THz electric field $\vec E(t,\vec r)$ to couple and drive the vortex via $ - \vec{E} \cdot \vec{P}$.
 In the continuum limit, the net ferroelectric polarization is
%
$\vec{P}=\xi_{m}\left[ (\vec{M} \cdot \vec{\nabla}) \vec{M}-\vec{M}(\vec{\nabla} \cdot \vec{M}) \right]$  \cite{KatsuraH:2005,Mostovoy:2006,JiaC1:2006},
%
where $\vec{M}$ is the saturation magnetization, and $\xi_{m}$ is the ME coupling constant.   Thus, the additional ME-driven effective field acting on the spin system reads
\begin{equation}
\vec{H}_{me}=2 \xi_{m} \left[ \vec{E} \cdot ( \vec{\nabla} \vec{M}) - (\vec{\nabla} \cdot \vec{M}) \vec{E} \right] + \xi_{m} \vec{M} \times (\vec{\nabla} \times \vec{E}).
\end{equation}
%
 For an electric field normal to the magnetic plane (hereafter defined by the $\vec{e}_{z}$ direction), the first term in $\vec{H}_{me}$  acts on the magnetization as the field due to an  interfacial Dzyaloshinskii-Moriya interaction (DMI) \cite{Bogdanov:2001hr,{Fert:2013fq}},
\begin{equation}
\vec{H}_{DM}^{\rm me}= 2 \xi_{m} E_{z} \left( \frac{\partial M_{z}}{\partial x}, \frac{\partial M_{z}}{\partial y},
- \frac{\partial M_{x}}{\partial x}-\frac{\partial M_{y}}{\partial y} \right).
\end{equation}
The curl of the electric field entering  the last term of  $\vec{H}_{me}$ is expressible  via Maxwell's equation as  $\vec{\nabla} \times \vec{E} = - \frac{\partial \vec{B}}{\partial t} = -  \mu_{0} (1+ \chi_{m}^{-1}) \frac{\partial \vec{M}}{\partial t}$ (where $\chi_{m}$  is the material magnetic susceptibility \cite{Hickey:2009hb})  pointing to an additional route for magnetic energy transfer and  precessional damping control. Thus, in principle  it is possible to  realize all-electric magnetization switching through ME coupling.
%

\begin{figure*}[t]
\centering
    \includegraphics[scale =1.2]{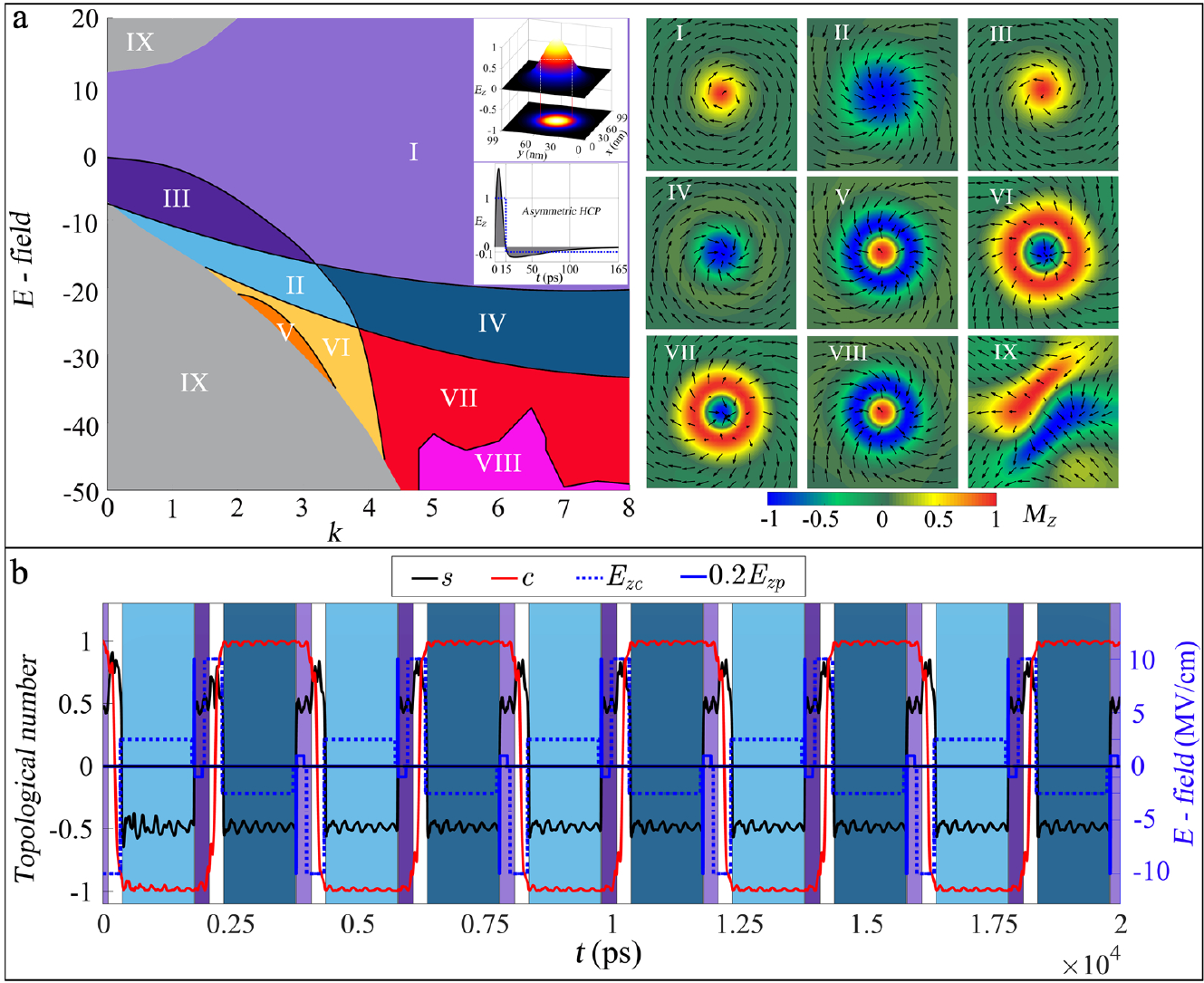}
     \caption{\textbf{Electric field control of magnetic vortex.} (\textbf{a}) Phase diagrams of remanent magnetization configurations (I-IX) reached by applying static Gaussian electric field to the initial vortex state I. (\textbf{b}) Reproducible switching between four magnetic vortex states: I $\rightarrow$ II $\rightarrow$ III $\rightarrow$ IV $\rightarrow$ I ... ... (labeled respectively by the corresponding aided colors in \textbf{a}) by a simple sequence of two  time asymmetric pulses (also referred to as half cycle pulse, HCP), $E_{zp}$ (dotted blue line) and $E_{zc}$ (solid blue line). Other remaining white areas denote the transition processes without well-defined vortex states. $s$ (black solid  line) and $c$ (red solid line) are the skyrmion and chirality numbers of magnetization configurations, respectively. \textbf{Inset}: the above panel shows  the diffraction-limited  spatial intensity Gaussian profile distribution of  the employed electric field pulses; the black envelop in the bottom panel shows one cycle of the  time profile of an experimentally feasible strongly asymmetric pulse;  the blue dotted line is the numerical shape used in  simulations. In both cases, the time integral over  pulse duration is zero,  as required by the Maxwell equation. }
    \label{Fig::vortex}
\end{figure*}
%

In fact, as demonstrated here,  the ME coupling allows for a remarkably swift, detailed and controlled  manipulation of  the magnetic vortex configurations by electric means.
 In addition to achieving a precisely and independently reproducible reversal of vortex  polarity and/or chirality by a simple sequence of feasible asymmetric THz pulses (cf. the inset of Fig.\ref{Fig::vortex}a) \cite{Keren-Zur:2019,Lee:2017,Hwang:2015,Lepeshov:2017,Welsh:2014,Jamal:2017}, we observe a new phenomenon, namely a packaged-skyrmion mediated picosecond switching of magnetic vortex. This switching does not involve the gyrotropic motion with the creation and subsequent annihilation of a vortex-antivortex pair \cite{B. Van:2006}. In addition to the technological relevance, the  results  provide a deeper insight into the fundamentals  of the ME effects and magnetization relaxation dynamics. Furthermore, the findings point to exciting  opportunities for exploiting the impressive recent advances in  THz sources  \cite{Keren-Zur:2019,Lee:2017,Hwang:2015,Lepeshov:2017,Welsh:2014} for  spintronics devices.

\section{Phase diagrams}

\begin{figure*}[t]
  \centering
      \includegraphics[scale =1.2]{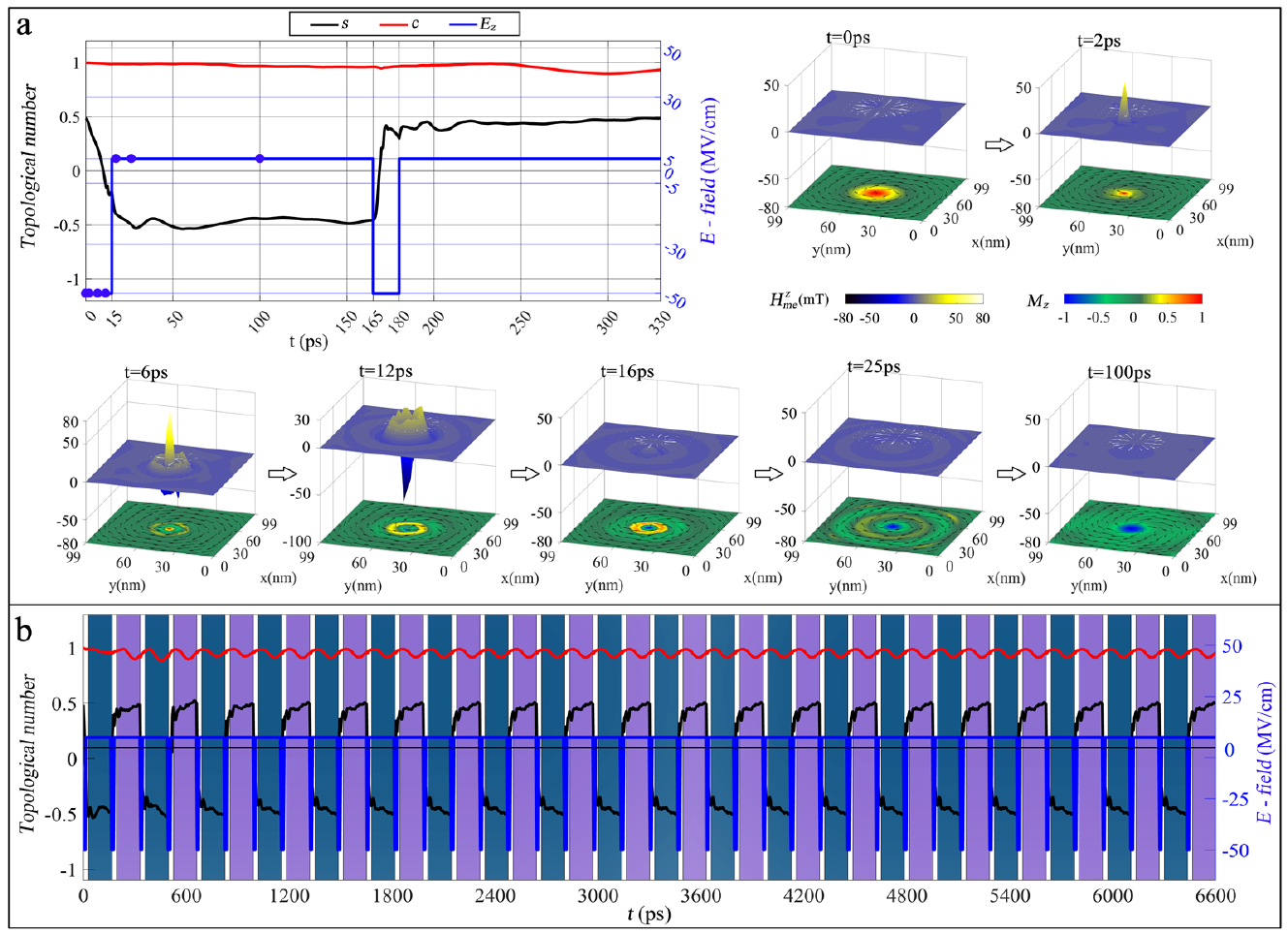}
       \caption{\textbf{Ultrafast vortex core switching with strongly  asymmetric time envelope} ($k=6$). (\textbf{a}) Time-resolved micromagnetic simulations showing the dynamic configurations of magnetization and perpendicular effective ME fields.  The dynamic process of the vortex switching can be effectively described by the time-evolution of the skyrmion number ($s$, black solid line), and the chirality number ($c$, red solid line). (\textbf{b}) Reproducible control of the vortex polarity. Two vortex states (I and IV) having opposite core polarization are labelled with the same color scheme used in Fig.\ref{Fig::vortex}.}
      \label{Fig::core}
  \end{figure*}

  \begin{figure*}[t]
    \centering
        \includegraphics[scale =1.2]{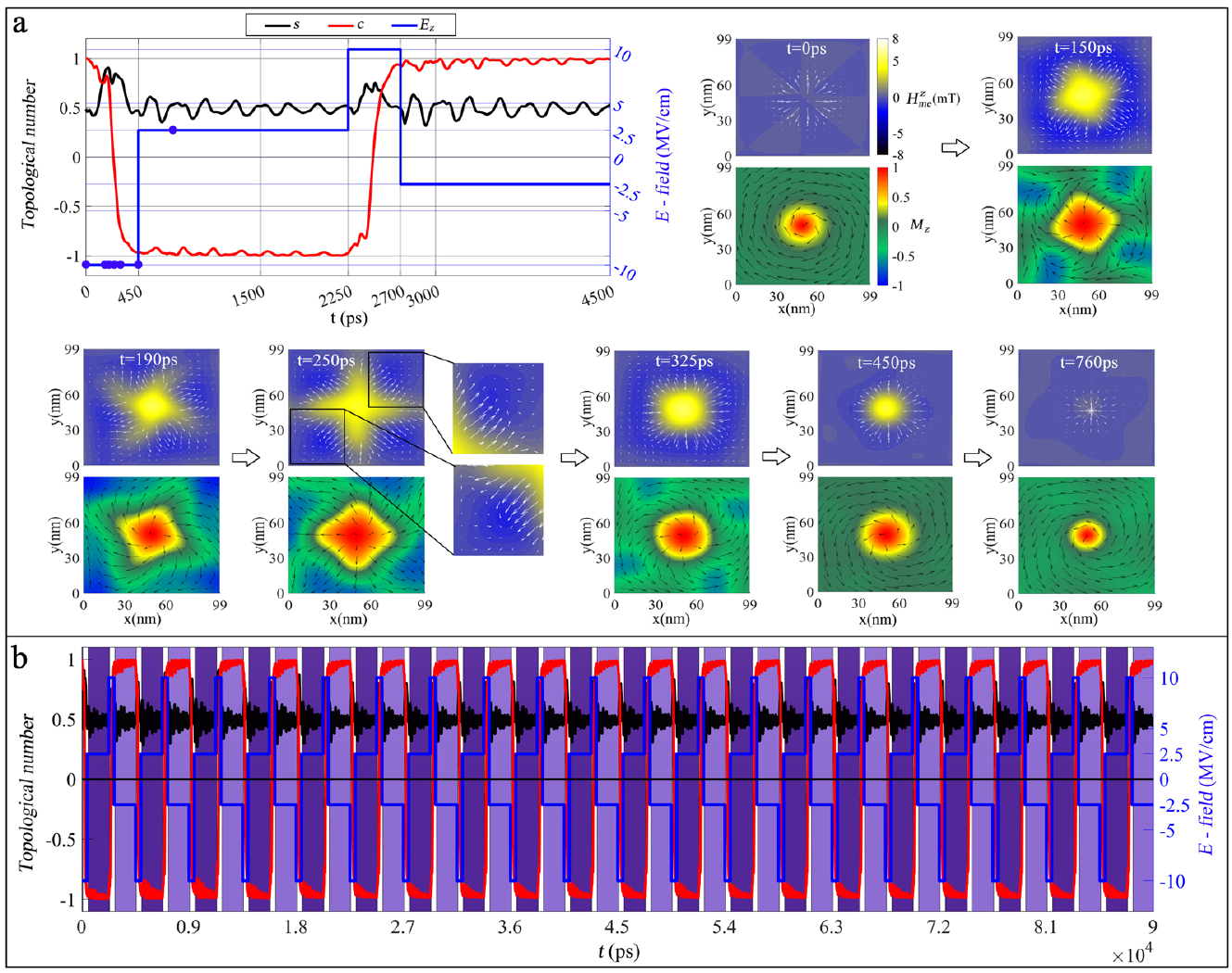}
         \caption{\textbf{Vortex chirality reversal.} (\textbf{a}) Time evolution of the skymion and chirality numbers driven by the Gaussian pulse ($k=1$) sequences with opposite field direction. Snapshot images illustrating in-plane ME field (above) and magnetization configuration (below) at different stages of the vortex dynamics. (\textbf{b}) Reliable chirality switching between the stable vortex state I ($s=1/2$ and $c=1$) and III ($s=1/2$ and $c=-1$) by HCPs same as in (\textbf{a}) .}
        \label{Fig::chirality}
    \end{figure*}
    \begin{figure*}[t]
    \centering
        \includegraphics[scale =1.8]{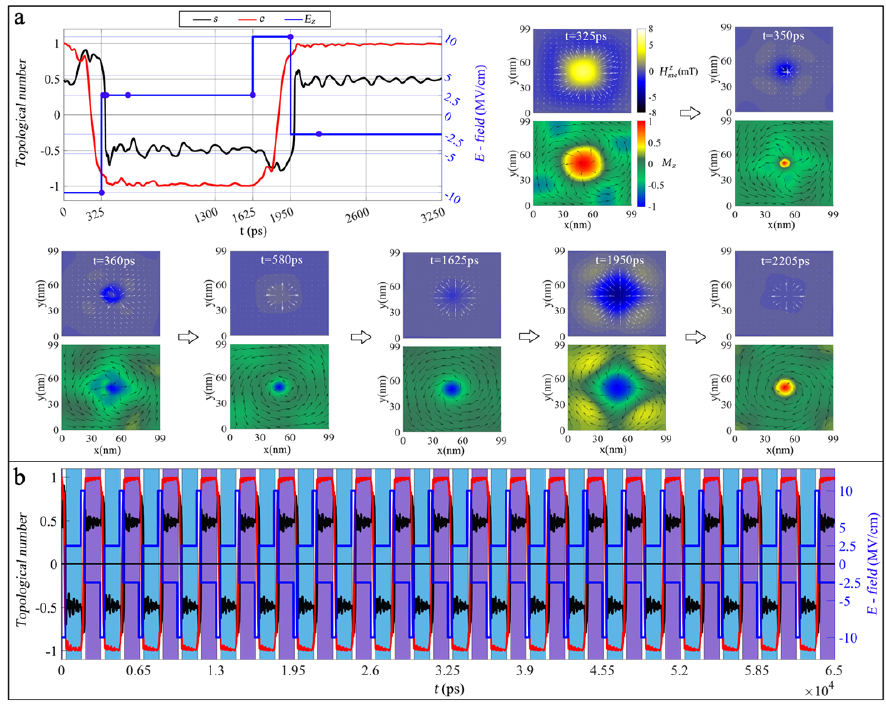}
         \caption{\textbf{Simultaneous control of vortex polarity and chirality.} (\textbf{a}) Relative shorter HCPs compared that in the Fig.\ref{Fig::chirality} is used to reverse the vortex polarity and chirality simultaneously. Vortex dynamics before $325$ ps are the same as shown in Fig.\ref{Fig::chirality}a, different and important behavior are illustrated in the snapshots after $325$ ps. (\textbf{b}) Vortex switching between the vortex state I ($s=1/2$ and $c=1$), and the vortex state II (with opposite vortex polarity and chirality, i.e., $s=-1/2$ and $c=-1$).}
        \label{Fig::pc}
    \end{figure*}

The magnetization dynamics of the vortex is investigated by the finite-difference micromagnetic simulations with GPU acceleration based on the Landau-Lifshitz-Gilbert (LLG) equation,
\begin{equation}
\begin{aligned}
\frac{d\vec{M}}{dt}=&-\frac{\gamma}{1+\alpha^2}(\vec{M}\times\mu_0\vec{H}_{eff})\\
     &-\frac{\alpha\gamma}{(1+\alpha^2)M_s}[\vec{M}\times(\vec{M}\times\mu_0\vec{H}_{eff})],
\end{aligned}
\end{equation}
where $\gamma$ is the gyromagnetic ratio and $\vec{H}_{eff}$ is the effective magnetic field, which includes the isotropic Heisenberg exchange field, the uniaxial magneto-crystalline anisotropy field, the magnetostatic demagnetising field, and the ME field $\vec{H}_{me}$. In the present study, the material parameters are chosen for permalloy (Py, a magnetically very soft material of great technological importance) at room-temperature: the exchange constant $A=13$ pJ/m, the saturation magnetization $M_s=8 \times10^5$ A/m, and the magneto-crystalline anisotropy $K_{u} =0$. The Gilbert damping constant is set to $\alpha=0.05$, which is relatively larger than the value of a  bulk system but is in a reasonable range for a Py thin film  \cite{Tserkovnyak:2002}. The ME coupling strength $\xi_{m}=1$ pC/m \cite{Wang:2019gb}. The initial stable magnetic vortex with the positive polarity ($p=1$) and clockwise chirality ($c=1$) is prepared in a typical Py square (width of $99$ nm and thickness of $3$ nm) without any external magnetic/electric fields (cf. the magnetization configuration I in the Fig.\ref{Fig::vortex}a). By sequentially applying two types of time-asymmetric electric field pulses with different amplitude and pulse duration to the initial state I, we observe an  ultrafast and highly reproducible switching between four vortex states, as demonstrated in Fig.\ref{Fig::vortex}b (cf. also Supplementary Video 1). The time required for the core polarization reversal is of the order of $15$ ps and the chirality reversal is around $325$ ps. These are very high speeds for a magnetic vortex switching, faster than any previously reported speeds achieved by other control schemes \cite{Zheng:2017bq}. In addition, the  reliable repeatability of the switching process is a clear advantage.

For a better understanding  let us  at first refine the possible stable magnetization configurations under time-independent but spatially  inhomogeneous electric fields, meaning  a  Gaussian profile of a  static electric field applied along the $z$ axis
\begin{equation}
\vec{E}(\vec{r}) = [0, 0, E_{z} e^{-k(x^{2}+y^{2})/R^{2}}].
\end{equation}
Here $R=51$ nm and the width of Gaussian distribution is determined by the parameter $k$.  Perturbing with  $ \vec{E}(\vec{r})$ the prepared vortex state I with $p=1$ and $c=1$, the stable topological magnetic configurations can be characterized by the skyrmion number $s$ and the chirality number $c$ as
\begin{equation}
s = \sum_{\vec{r}} \chi_{\vec{r}} ~~\text{and}~ c= \frac{1}{2\pi}\int_0^{2\pi} \sin(\phi-\theta) d\theta
\end{equation}
where $\phi=\arctan(\vec{M}_{{x}}/\vec{M}_{{y}})$ and $\chi_{\vec{r}}=\vec{M}_{\vec{r}}\cdot(\vec{M}_{\vec{r}+\vec{x}}\times\vec{M}_{\vec{r}+\vec{y}})+\vec{M}_{\vec{r}}\cdot(\vec{M}_{\vec{r}-\vec{x}}\times\vec{M}_{\vec{r}-\vec{y}})$. For example, a vortex with a winding number $n=1$ and core polarization $p$ has a half-integer skyrmion number $s=np/2$, while an ideal skyrmion (with an oppositely  pointing direction of spins at $r \rightarrow 0$  and $r \rightarrow \infty$, respectively) has integer $s= np$  \cite{Tretiakov:2007,Nagaosa:2013cc}. In Fig.\ref{Fig::vortex}a, we present a more detailed examination of the relaxation into remanent states from the initial configuration I under different Gaussian electric fields. In addition to the four desired vortex states (I-IV) and one complex distorted state IX, four new packaged skyrmion-like stable configurations (V-VIII) emerge in the presence of large and narrow Gaussian electric fields. The skyrmion number of these four new, well-shaped  configurations is less than $0.5$ and mostly around $0.3$ by integrating $\chi_{\vec{r}}$ over the whole square, but it may approach $\pm 1$ if the radius of the integral area is shrunken to be around 0.7 halfwidth of the corresponding Gaussian field. From the topological point of view, these packaged skyrmions with localized topological charge $|s| =1$ are very important for a switching between the $s = \pm 1/2$ topological sectors, as shown in the following figures,  they indeed occur dynamically during the vortex switching processes, assisting the polarity and chirality reversal.

The phase diagrams suggest the following possible vortex transitions: (i) core polarization reversal from the state I ($s = 1/2$ and $c=1$) to IV ($s=-1/2$ and $c=1$), (ii) chirality reversal from the state I ($s = 1/2$ and $c=1$) to III ($s=1/2$ and $c=-1$), and (iii) simultaneous polarity and chirality reversal from the state I ($s = 1/2$ and $c=1$) to II ($s=-1/2$ and $c= -1$) by applying different amplitude ($E_{z}$) and halfwidth ($k$) electric fields. However, it should be noted that the dynamic switching processes are inessential. The transition conditions in the phase diagrams are mainly  determined by  external parameters: the pulse amplitude and the pulse duration.

\section{Vortex switching}

\textbf{Vortex core switching.} The detail of the magnetization dynamics of the vortex core reversal processes, including the time evolution of the ME field, are demonstrated in Fig.\ref{Fig::core}a (see also Supplementary Video 2).
The pulses we are employing are generally feasible. In fact much shorter and stronger pulses were reported (for instance \cite{Goul}).  The  pulse consists of one strong and short part (we call head) and much longer and weaker part with opposite polarity (we call tail).  As shown in full details in \cite{Jamal:2017}, as long as the duration of the head is shorter than the typical time scale of the system (here the precessional periode), the exact shape of the pulse is subsidiary.

In the typical case shown in Fig.\ref{Fig::core}a we observe,
as the strong ($50$ MV/cm) and short ($15$ ps) head part of the asymmetric HCP ($k=6$) reaches  the initial vortex I, a  ME field with positive core but negative circling is dynamically induced. With  time passing, this effective magnetic field reverses its direction and the temporal packaged skyrmion (VII) develops reaching full formation at around $14$ ps (before the pulse tail). Eventually the packaged skyrmion decays into spin waves and the vortex IV with opposite polarity is stabilized during the weak ($5$ MV/cm) and long ($150$ ps) tail of the HCP.  The vortex IV can be switched back to the vortex I by a second sequence (same-type) pulse following the very similar processes, IV $\xrightarrow{\text{VIII}}$ I with the emission of spin waves. These vortex polarity reversals show very high and reliable reproducibility, as demonstrated in Fig.\ref{Fig::core}b and Supplementary Video 3. Clearly, the underlying mechanism is different from the usual gyrotropic core excitation, in which the vortex core switching dynamics involves the creation and annihilation of a vortex-antivortex pair, accompanying with the injection of  a magnetic monopole \cite{B. Van:2006,R. Hertel:2007,Pigeau:2010fy,Thiaville:2003}. Here, the change of the topological charge $\Delta s = \pm 1$ between the two vortex sections is carried by the formation of the  intermediate packaged skyrmion with highly-localized skyrmion number $s= \pm 1$. It is worth emphasizing  that the core reversal speed can be enhanced further by a stronger and/or shorter time-asymmetric electric field pulse.  This is particularly useful from the application point of view, as short pulses cause less material damage.

\textbf{Vortex chirality reversal.} In terms of applications, the chirality switching is highly desirable for storing two bits of digital information in magnetic vortex-based devices. Chirality is akin to symmetry and  can be changed in many cases by introducing an asymmetry in ferromagnetic disks, such as masked disk \cite{Gaididei:2008ez}, odd-side regular polygonal nano-magnets \cite{Yakata:2010ka}, and rectangular thin films with the aspect ratio $2:1$ \cite{Li:2018gb}. However, such methods cannot be applied to a symmetrical square disk with even number of sides due to symmetrical nucleation energy during vortex formation. S. Yakata et al. proposed  an alternative method for controlling the vortex chirality in squared permalloy dot by the circular Oersted field \cite{Yakata:2011eh}. Here, we show that the vortex chirality reversal can be simply realized by  
using time-asymmetric electric field pulses as well. Compared with the vortex core reversal, the chirality switching needs relatively weaker and longer HCPs with small $k$.  Fig.\ref{Fig::chirality} and the Supplementary Video 4 evidence the only-chirality-switching via  HCPs  with peak  amplitude $10$ MV/cm and duration of $450$ ps. The slower part of the HCP peaks at $2.5$ MV/cm and has a duration of $1800$ ps. A short time after the excitation of the vortex I, a hedgehog skyrmion without favorite chirality develops, and the skyrmion number reaches its maximum $s \approx 1$ around $190$ ps.
This unstable (global) N\'{e}el-type skyrmion relaxes then into the vortex state III before the end of HCP's head (450 ps), at which the vortex chirality has been completely reversed. Such vortex state (III) survives  the pulse tail and becomes more stable through the emission of spin waves. Similar to the vortex core switching, well-defined reproducible performance of chirality switching can be realized by two inversion recovery pulse sequences (see Fig.\ref{Fig::chirality}b and also Supplementary Video 5).

\textbf{Simultaneous control of vortex polarity and chirality.}
As demonstrated in Fig.\ref{Fig::pc}b and Supplementary Video 6, we find that by just slightly shortening  the pulse duration to terminate the relaxation from the hedgehog skyrmion to the vortex state III (cf. Fig.\ref{Fig::chirality}a), the skyrmion  dissolves completely after the peak pulse.
 The reversed chirality, however, persists into the pulse tail, and the magnetic core is reversed as well pretty soon after the chirality reversal. The vortex configuration II emerges and is then stabilized by releasing  spin waves. The simultaneous control of the vortex core and chirality show a  highly reliable  repeatability, I $\rightarrow$ II $\rightarrow$ I ... ... (cf Fig.\ref{Fig::pc}b and also Supplementary Video 7).

Furthermore, it is also possible to obtain independent switching between two random vortex states.  Efficient and reliable manipulation of four magnetic vortex states is achievable by any two combinations of the above three switching procedures. 
For instance as shown in Fig.\ref{Fig::vortex}b, the vortex switchings contains totally two predictable dynamic procedure, I $\rightarrow$ II $\rightarrow$ III $\rightarrow$ IV $\rightarrow$ I ... ... However, it should be noted that, in order to ensure a more stable hybrid modulation of the vortex, a $42.5$ ps interval is set between each pulse.

\section{Conclusions}
In the quest for an ultrafast optical switching of magnetic order (\cite{Kimel:2019,Kampfrath:2011}, and references therein), THz fields  such as those employed here were demonstrated to be
a powerful tool  \cite{Schlauderer:2019,Yang:2017}. Pulsed THz drivings  allow also for new schemes of  non-invasive non-processional switching  \cite{Sukhov:2009,Yamaguchi:2010}.
 The fantastic advances in generating and controlling such pulses in the near fields
  (for instance via plasmonic structures) \cite{Keren-Zur:2019,Lee:2017,Hwang:2015,Lepeshov:2017,Welsh:2014} call for exploiting them for data writing/deleting and moving of bits stored in topological magnetic excitations such as magnetic vortices. The intrinsic magnetoelectric interaction that stems from the spin-noncollinearity in magnetic vortices serves for coupling to the electric field component of the THz pulse. In our  simulations we used a sequence of time asymmetric pulses.  The actual shape of the pulse is, however, not decisive, only the time asymmetry and peak field value do matter \cite{Jamal:2017}. The time asymmetry is related to the duration of the positive to the negative polarity parts of the pulse. Hence, the phenomena predicted here are robust to reasonable variations in the pulse characteristics ensuring so a reliable operation of the vortex-based data storage. In addition to this advantageous feature,  the reproductivity of the picosecond  switching of magnetic vortex  polarity and chirality by a simple sequence the THz pulses are also useful. The topological protection and the speed restriction on the vortex-core switching can be lifted when forming and utilizing packaged skyrmion. Unlike  gyrotropic excitation, the packaged-skyrmion driven switching of the magnetic vortex does not involve displacement of the vortex core, which bears a great advantage for ultrafast magnetic recording. Spatially localized THz fields, as realized in near-field optics are advantageous for vortex core reversal. This points to using plasmonic near fields for vortex steering. Far fields are also useful for vortex chirality control but are diffraction limited (as also assumed in this work). It is however possible to go beyond the diffraction limit by using a normal metal mask or SNOM techniques \cite{Bazylewski:2017fz}. The independent reversal of vortex polarity and chirality, presented here, demonstrate a straightforward  mean for the  electric field control of the vortex states. Thus,  the current findings for the dynamic   magnetic vortex switching are of direct and significant relevance  for  optical-based ultrafast sprintronics .

\section*{Supplementary material}
{See supplementary material for the detailed videos of the magnetic vortex switching processes.}

\begin{acknowledgments}
{
This work is supported by the National Natural Science Foundation of China (Nos. 11474138 and 11834005), the German Research Foundation (Nos. SFB 762 and SFB TRR227), the Program for Changjiang Scholars and Innovative Research Team in University (No. IRT-16R35), and the Ministry of Science and Technology of China through grants CN-SK-8-4.
}
\end{acknowledgments}

\end{document}